\documentclass[prl,aps,twocolumn]{revtex4}
\usepackage{amsfonts,amssymb,amsmath,amsthm,mathrsfs}
\usepackage{bbold,mathtools}
\usepackage{hyperref} 
\usepackage{xcolor}

\newcommand{\bi}[2]{\binom{#1}{#2}}

\def\>{\rangle}
\def\<{\langle}

\def\floor#1{{\left \lfloor {#1} \right \rfloor  }}

\begin{document} 

\title{Permutation-invariant codes encoding more than one qubit}%

\author{Yingkai \surname{Ouyang}  }
\affiliation{Singapore University of Technology and Design, 8 Somapah Road, Singapore}
\email{yingkai\_ouyang@sutd.edu.sg}

\author{Joseph \surname{Fitzsimons}  }
\affiliation{Singapore University of Technology and Design, 8 Somapah Road, Singapore}
\affiliation{Centre for Quantum Technologies, National University of Singapore, 3 Science Drive 2, Singapore}

\begin{abstract}
A permutation-invariant code on $m$ qubits is a subspace of the symmetric subspace of the $m$ qubits.
We derive permutation-invariant codes that can encode an increasing amount of quantum information while suppressing leading order spontaneous decay errors.
To prove the result, we use elementary number theory with prior theory on permutation invariant codes and quantum error correction.
\end{abstract}

\maketitle

The promise offered by the fields of quantum cryptography \cite{BB84,Eke91} and quantum computation \cite{nielsen-chuang} has fueled recent interest in quantum technologies.
To implement such technologies, one needs a way to reliably transmit quantum information,
which is inherently fragile and often decoheres because of unwanted physical interactions.
If a decoherence-free subspace (DFS) \cite{ZaR97} of such interactions were to exist, 
encoding within it would guarantee the integrity of the quantum information.
Indeed, in the case of the spurious exchange couplings \cite{Blundell}, 
the corresponding DFS is just the symmetric subspace of the underlying qubits.
In practice, only approximate DFSs are accessible because of small unpredictable perturbations to the dominant physical interaction \cite{LBW99}, and
using approximate DFSs necessitate a small amount of error correction.
When the approximate DFS is the symmetric subspace, permutation-invariant codes can be used to negate the aforementioned errors \cite{Rus00,PoR04,ouyang2014permutation}.
However, as far as we know, all previous permutation-invariant codes encode only one logical qubit \cite{Rus00,PoR04,ouyang2014permutation}.
One may then wonder if there exist permutation-invariant codes
that can encode strictly more quantum information than a single qubit whilst retaining some capability to be error-corrected.

The first example of a permutation-invariant code which encodes one qubit into 9-qubits while being able to correct any single qubit error was given by Ruskai over a decade ago \cite{Rus00}. 
A few years later, Ruskai and Pollatshek found 7-qubit permutation invariant codes encoding a single qubit which correct arbitrary single qubit errors \cite{PoR04}.
Recently permutation-invariant codes encoding a single qubit into $(2t+1)^2$ qubits that correct arbitrary $t$-qubit errors has been found \cite{ouyang2014permutation}. Here, we extend the theory of permutation-invariant codes. 
Our permutation-invariant code $\mathcal C$ has 
as its basis vectors the logical 1 of $D$ distinct permutation invariant codes given by \cite{ouyang2014permutation}, where each such code encodes only a single qubit.
Surprisingly, this simple construction can yield a permutation-invariant code encoding more than a single qubit while correcting spontaneous decay errors to leading order.

Permutation-invariant codes are particularly useful in correcting errors induced by {\em quantum permutation channels with spontaneous decay errors}, with Kraus decomposition 
$\mathcal N(\rho) = \mathcal A ( \mathcal P ( \rho) )  =
\sum_{\alpha, \beta} A_\beta P_\alpha \rho P_\alpha ^\dagger A_\beta$,
where $\mathcal P$ and $\mathcal A$ are quantum channels satisfying the completeness relation
$\sum_{\alpha} P_\alpha  ^\dagger P_\alpha 
= \sum_{\beta} A_\beta ^\dagger A_\beta = \mathbb 1 $ and $\mathbb 1$ is the identity operator on $m$ qubits.
The channel $\mathcal P$ has each of its Kraus operators $P_\alpha$ proportional to $e^{i \theta_\alpha \hat a_\alpha}$, where $\theta_\alpha$ is the infinitesimal parameter and the infinitesimal generator $\hat a_\alpha$ is any linear combination of exchange operators. 
By a judicious choice of $\theta_\alpha$ and $\hat a_\alpha$, the channel $\mathcal P$ can model the stochastic reordering and coherent exchange of quantum packets as well as out-of-order delivery of classical packets \cite{Pax97}.
The channel $\mathcal A$ on the other hand models spontaneous decay errors, otherwise also known as amplitude damping errors, where an excited state in each qubit independently relaxes to the ground state with probability $\gamma$.
Our permutation-invariant code is inherently robust against the effects of channel $\mathcal P$, and can suppress all errors of order $\gamma$ introduced by channel $\mathcal A$, and is hence approximately robust against the composite noisy permutation channel $\mathcal N$.

We quantify the error correction capabilities of our permutation-invariant codes $\mathcal C$ with code projector $\Pi$ beginning from the approximate quantum error correction criterion of Leung {\em et\ al.} \cite{LNCY97}. 
Since the Kraus operators $P_\alpha$ of the permutation channel leave the codespace of any permutation-invariant code unchanged, it suffices only to consider the effects of the amplitude damping channel $\mathcal A$.
The optimal entanglement fidelity between an adversarially chosen state $\rho$ in the permutation-invariant codespace and error-corrected noisy counterpart is just
\begin{align}
1 -\epsilon = \sup_{\mathcal R} \inf_{\rho} \mathcal F_e (\rho, \mathcal R \circ \mathcal A),
\label{eq:eps-def}
\end{align}
where $\epsilon$ is the the {\em worst case error} \cite{ouyang2014permutation} that we need to suppress.
Lower bounds for the above quantity can be found using various techniques from the theory of optimal recovery channels \cite{BaK02,Fletcher08,Yam09,Tys10,BeO10,BeO11,ouyang2014permutation}, 
but we restrict our attention to the simpler (but suboptimal) approach of \cite{LNCY97,ouyang2014permutation}.
Suppose that we can find a truncated Kraus set $\Omega$ \cite{ouyang2013truncated} of the channel $\mathcal A$ such that for every distinct pair of $A,B \in \Omega$, the spaces $A \mathcal C$ and $B \mathcal C$ are pairwise orthogonal.
Then the truncated recovery map of Leung {\em et\ al.} 
$\mathcal R_{\Omega,\mathcal C} (\mu):= \sum_{A \in \Omega} \Pi U_A ^\dagger \mu U_A \Pi $ is a valid quantum operation, where $U_A$ is the unitary in the polar decomposition of $A \Pi = U_A \sqrt{ \Pi A ^\dagger A \Pi }$. Since $\mathcal R_{\Omega,\mathcal C} $ is now a special instance of a recovery channel in Eq.~(\ref{eq:eps-def}), we trivially get
$\epsilon \le 1- \inf_{\rho} \mathcal F_e (\rho, \mathcal R_{\Omega,\mathcal C} \circ \mathcal A).$
As explained in \cite{ouyang2014permutation}, the analysis of Leung {\em et\ al.} \cite{LNCY97}  allows one to show that
\begin{align}
\mathcal F_e (\rho, \mathcal R_{\Omega,\mathcal C} \circ \mathcal A) 
\ge \sum_{A \in \Omega}\lambda_A,
\end{align}  
where
$\lambda_A= \min_{\substack{|\psi\> \in \mathcal C \\ \<\psi|\psi\> = 1 }} 
\<\psi | A ^\dagger A |\psi\>$ quantifies the worst case deformation of each corrupted codespace $A \mathcal C $. 

The symmetric subspace of $m$ qubits is central to the study of permutation-invariant codes, 
and has a convenient choice of basis vectors, namely the {\em Dicke states} \cite{BGu13,MHT12,TGG09,ouyang2014permutation}. A Dicke state of weight $w$, denoted as 
$|{\rm D}^m_w\>$, is a normalized permutation-invariant state on $m$ qubits with a single excitation on $w$ qubits. 
Our code $\mathcal C$ is the span of the logical states $|d_L\>$ for $d = 1,\dots, D$, and these states can be written as superposition over Dicke states, with amplitudes proportional to the square root of the binomial distribution. 
Namely for positive integers $n_d$ and $g_d$,
\begin{align}
|d_L\> = 
\sum_{j \in \mathcal I_d }  \sqrt{ \frac{\bi{ n_d}{j}}{2^{n_d-1}}} | {\rm D}_{g_d j}^m\>
\label{eq:new-pi-states}
\end{align}
and the set $\mathcal I_d$ comprises of the odd integers 
from 1 to $2 \floor{\frac{n_d-1}{2}}+1$..
The states $|d_L\>, A |d_L\>$ can be made to be pairwise orthogonal 
via a judicious choice of constraints on the positive integer parameters $n_1, \dots, n_D$, $g_1, \dots g_D$ and $m$.

We elucidate the case for $D \ge 3$ since permutation invariant codes encoding only one qubit \cite{ouyang2014permutation} are already known.
Here, we require $n_1, \dots, n_D$ to be pairwise coprime integers with $n_1 \le \dots \le n_D$, and define their product to be $N = n_1 \dots n_D$.
The length of our code is a polynomial in $N$, given by $m = N^q$ for any integer $q \ge 3$.
Moreover we set $g_d = N/n_d$ so that for distinct $d$ and $d'$, the greatest common divisor of $g_d$ and $g_{d'}$ is precisely gcd$(g_d, g_{d'}) = N/(n_d n_{d'}) > 1$, so that $g_d$ and $g_{d'}$ are not coprime.
Furthermore, we require that $g_d \ge 3 $, $n_d \ge 4$.

The reason for requiring $g_d$ and $g_{d'}$ to not be coprime is that it allows the inner products $\< d_L | d'_L\>$ and 
$\< d_L | A^\dagger	 B |d'_L\>$ to be identically zero for distinct $d$ and $d'$ and for any operators $A,B$ acting nontrivially on strictly less than $\frac{\min_d g_d}{2}$ qubits when $N$ is even. To see this, we analyze the linear Diophantine equation 
\begin{align}
x_{d,d'} g_d = y_{d,d'} g_{d'} + s \label{eq:linear-Diophantine},
\end{align}
with $s=0,\pm 1 $.
This linear Diophantine equation has a solution $(x_{d,d'},y_{d,d'})$ if and only if $s$ is a multiple of gcd($g_d,g_{d'}$).
Having gcd($g_d,g_{d'})> 1$ ensures that 
Eq.~(\ref{eq:linear-Diophantine}) has no solution for non-zero $s$ such that $|s| < {\rm gcd}(g_d,g_{d'})$.
When $s=0$, integer solutions $(x_{d,d'},y_{d,d'})$ where $0 < x_{d,d'} g_d = y_{d,d'} g_{d'} < N$ do not exist. 
To see this, note that the minimum positive solutions of Eq.~(\ref{eq:linear-Diophantine}) are precisely $x_{d,d'} = \frac{g_{d'}}{{\rm gcd}(g_d,g_{d'})}$ and 
$y_{d,d'} = \frac{g_d}{{\rm gcd}(g_d,g_{d'})}$, and hence
we must require that $\frac{g_d g_{d'}}{{\rm gcd}(g_d,g_{d'})} < N$ be an invalid inequality.
But our construction gives $\frac{g_d g_{d'}}{{\rm gcd}(g_d,g_{d'})} = 
\frac{g_d g_{d'} n_d n_{d'}}{N} = N$. 
This immediately implies several orthogonality conditions on the states given by 
Eq.~(\ref{eq:new-pi-states}) for large $n_1$.

We use a sequence of large consecutive primes and an even number to construct our sequence of coprimes.
We let $n_1 = p_k$, where $p_k$ denotes the $k$-th prime, 
and let $n_2 =  n_1+1$. We also let $n_j = p_{k+j-2}$ for all $j = 3, \dots, D$, which gives us our $D$ coprime integers.
The length of our code is $m =((p_k+1)(p_k \dots p_{k+D-2} ))^q$. 
In the special case when $D=3$, we can use the existence of twin primes $n_1$ and $n_3$ a bounded distance apart \cite{zhang2014bounded} (at most 600 apart \cite{maynard2013small}), and let 
$n_2 = n_1 + 1$, which yields 
$m  = (n_1 n_3(n_1 + 1))^q$.

The oft used Kraus operators for an amplitude damping channel on a single qubit are
$A_0 = |0\>\<0|  + \sqrt{1 - \gamma} |1\>\<1| $ and $A_1 = \sqrt{\gamma} |0\>\<1|  $ respectively, with $\gamma$ modeling the probability for a transition from the excited $|1\>$ state to the ground state $|0\>$.
On $m$ qubits, the Kraus operators of the amplitude damping channel have a tensor product structure, given by 
$A_{x_1} \otimes \dots \otimes A_{x_m}$ where $x_1, \dots,x_m = 0,1$.
We focus our attention on the Kraus operators
$K_0 = A_0 ^{\otimes m}$, and $F_j$ which applies $A_1$ on the $j$-th qubit and applies $A_0$ everywhere else for $j =1,\dots, m$.
The choice of Kraus operators for a quantum channel is not unique, and we can equivalently consider a subset of the Kraus operators in a Fourier basis. 
Namely, for $\ell =1, \dots, m$, we define
$K_\ell  = \frac{1}{\sqrt{m}} \sum_{j=1}^m \omega^{(\ell-1) (j-1) } F_j,$
where $\omega = e^{2\pi i /m }$.
We choose the set of Kraus operators that we wish to correct to be 
$\Omega =\{ K_0 , K_1, \dots, K_m  \}$.

Now the spaces $A \mathcal C$ and $B \mathcal C$ are orthogonal for distinct $A,B \in \Omega$. Note that for $\ell, \ell' = 1,\dots, m$,
\begin{align}
&\<d_L | K_\ell	^\dagger K_{\ell'} |d_L\>\notag\\
=&
\frac{1 }{m}\sum_{j=1}^m \sum_{j'=1}^m
\omega^{ -(\ell-1)(j-1) + (\ell' - 1)(j'-1) }
\<d_L | F_j	^\dagger F_{j'} |d_L\> \notag\\
=&
\sum_{j=1}^m \omega^{(\ell'-\ell)(j-1)} 
\<d_L | F_j	^\dagger F_{j} |d_L\> \notag\\
 +&
 \frac{1}{m} \sum_{d=1}^{m-1}  \sum_{j=1}^m 
\omega^{-(\ell - 1)(j-1) + (\ell'-1)(j-1+d)} 
\<d_L | F_j	^\dagger F_{j+d} |d_L\>, \label{eq:Kraus-fourier}
\end{align}
where the addition in the subscript is performed modulo $m$. Using the invariance of 
$\<d_L | F_j	^\dagger F_{j} |d_L\>$ and
$\<d_L | F_j	^\dagger F_{j'} |d_L\>$ 
for distinct $j,j' = 1,\dots, m$ along with the identity
\begin{align}
\sum_{d=1}^{m-1}  \sum_{j=1}^m 
\omega^{-(\ell - 1)(j-1) + (\ell'-1)(j-1+d)} 
&=
(m \delta_{\ell', 1} - 1) m \delta_{\ell,\ell'} \notag,
\end{align}
one can simplify (\ref{eq:Kraus-fourier}) to get
\begin{align}
&\<d_L | K_\ell	^\dagger K_{\ell'}  |d_L\>\notag\\
=&\delta_{\ell, \ell'} 
\left(
\<d_L | F_1	^\dagger F_1  |d_L\>
+
(m \delta_{\ell,1} -1) \<d_L | F_1	^\dagger F_m  |d_L\>
\right),  \label{eq:diagonalized-Krauses}
\end{align} 
which completes the proof of the orthogonality of $A \mathcal C$ and $B \mathcal C$ for distinct $A,B \in \Omega$.  

Now we have
\begin{align}
\<d_L| K_0 ^\dagger K_0 |d_L \>
&= \sum_{ t \in \mathcal I_d} \frac{ \bi{n_d}{t} } {2^{n_d-1}} (1-\gamma) ^{g_d t} 
  \notag\\
\<d_L| F_1 ^\dagger F_1 |d_L \>
&= \gamma\sum_{ t \in \mathcal I_d} \frac{ \bi{n_d}{t} } {2^{n_d-1}} (1-\gamma) ^{g_d t-1} 
\frac{g_d t}{m}
  \notag\\
\<d_L| F_1 ^\dagger F_m |d_L \>
&= \gamma\sum_{ t \in \mathcal I_d} \frac{ \bi{n_d}{t} } {2^{n_d-1}} (1-\gamma) ^{g_d t-1} 
\frac{g_d t(m-g_dt)}{m(m-1)}   .
\end{align}
Using the Taylor series 
$(1-\gamma)^{g_d t} = 1 - g_d t \gamma + \frac{g_d t(g_dt-1)}{2} \gamma^2 + O(\gamma^3)$ and 
$(1-\gamma)^{g_d t-1} = 1 - (g_d t-1) \gamma + O(\gamma^2)$ 
with the binomial identities 
 $\sum_{t = 0}^{n_d} t\bi{n_d} {t}  = 2^{n_d-1} n_d$,
$ \sum_{t = 0}^{n_d} t^2 \bi{n_d} {t}  = 2^{n_d-2} n_d ( n_d+1)$
and  
$ \sum_{t = 0}^{n_d} t^3 \bi{n_d} {t}  = 2^{n_d-3} n_d^2 ( n_d+3)$
  \cite{ouyang2014permutation,PBM86}, 
 we get
\begin{align}
\<d_L| K_0 ^\dagger K_0 |d_L \> 
&=
1-  \frac{N  }{2} \gamma  \notag\\
&\quad +   \left( \frac{N^2  +N g_d  }{8} - \frac{N}{4} \right) \gamma^2 + O(\gamma^3)   
\notag\\
\<d_L| F_1 ^\dagger F_1 |d_L \>
&= 
  \frac{N  }{2m} \gamma
-  \left(  \frac{N^2  +N g_d}{4m}  - \frac{N}{2m} \right) \gamma^2
\notag\\
&\quad +  O(\gamma^3)   
\notag\\
\<d_L| F_1 ^\dagger F_m |d_L \>
&= 
\frac{\left(    \frac{N  }{2} - \frac{N^2   + N g_d}{4m} \right) }{m-1}\gamma
\notag\\
&\quad +	  \frac{N^3  +3N^2 g_d}{8m(m-1)}  \gamma^2
\notag\\
& 
\quad	 -   \frac{ (N^2  +N g_d)\left( 1+ \frac{1}{m} \right)-2N}{4(m-1)}  \gamma^2
	 \notag\\
& \quad +  O(\gamma^3).
\end{align}
Now for all $|\psi\> \in \mathcal C$ where $\<\psi|\psi\> = 1$, we can write $|\psi\> = \sum_{d=1 }^D a_d |d_L\>$ such that $\sum_{d =1 }^D |a_d|^2 =  1 +
  O(2^{-n_1})  $ \footnote{The term $O(2^{-n_1}) $ arises because of the slight non-orthogonality of the states $|d_L\>$.}.
Hence for all $A \in \Omega$,
$\<\psi | A ^\dagger A | \psi\> = \sum_{d = 1 }^D |a_d|^2 \< d_L |  A ^\dagger A | d_L\>$ which implies that 
$\lambda_A \ge \min_{d=1, \dots, D} \< d_L  | A ^\dagger A | d_L \> 
{ (1+O(2^{-n_1}) )}$.
This implies that
\begin{align}
1- \epsilon
&\ge
 1 - \frac{ N g_1  }{ 4m } \gamma
 - \frac{c N^2}{8} \gamma^2  + O(\gamma^3) { + O(2^{-n_1})  },
\end{align} 
where
\begin{align}
c  =  1 +  \frac{2g_D-g_1}{N} - \frac{2}{N} + \frac{ 3g_1 }{m}   
 + \frac{4 g_1}{ N} .
\end{align}
Since $m = N^q$, $1-\epsilon \ge 1 - \frac{1}{4N^{q-2}} \gamma  - \frac{c N^2}{8} \gamma^2  + O(\gamma^3)  + O(2^{-n_1})  $ and for fixed $N$ and large $q$, the asymptotic error is second order in $\gamma$ with 
$\epsilon \sim \frac{ c'  N^2}{8} \gamma^2 + O(\gamma^3) {  + O(2^{-n_1})  }$, where $c' = 1 +  \frac{2g_D-g_1}{N} - \frac{2}{N} + \frac{4 g_1}{ N}$.

In summary, we have generalized the construction of permutation-invariant codes to enable the encoding of multiple qubits while suppressing leading order spontaneous decay errors. These permutation-invariant codes might allow for the construction of new schemes in physical systems, such as improved quantum communication along isotropic Heisenberg spin-chains \cite{BuB05PRA,BGB05,BuB05NJP,SJBB07}. Symmetry of error-correction codes have also recently been exploited to symmetrise prover strategies in the context of interactive proofs \cite{FV,Ji}, and so the extremely high symmetry of the codes studied here may also have theoretical implications.

This research was supported by the Singapore National Research Foundation under NRF Award No. NRF-NRFF2013-01.
Y. Ouyang also acknowledges support from the Ministry of Education, Singapore.

\bibliography{../../../mybib}{}

\begin{thebibliography}{10}

\bibitem{BB84}
C.~H. Bennett and G.~Brassard, ``{Quantum cryptography: Public key distribution
  and coin tossing},'' in {\em Proceedings of IEEE International Conference on
  Computers, Systems and Signal Processing}, vol.~175, New York, 1984.

\bibitem{Eke91}
A.~K. Ekert, ``Quantum cryptography based on {B}ell's theorem,'' {\em Phys.
  Rev. Lett.}, vol.~67, pp.~661--663, Aug 1991.

\bibitem{nielsen-chuang}
M.~A. Nielsen and I.~L. Chuang, {\em {Quantum Computation and Quantum
  Information}}.
\newblock Cambridge University Press, second~ed., 2000.

\bibitem{ZaR97}
P.~Zanardi and M.~Rasetti, ``{Noiseless Quantum Codes},'' {\em Phys. Rev.
  Lett.}, vol.~79, pp.~3306--3309, Oct. 1997.

\bibitem{Blundell}
S.~Blundell, {\em {Magnetism in Condensed Matter}}.
\newblock Great Clarendon Street, Oxford OX2 6DP: Oxford master series in
  condensed matter physics, first rnote~ed., 2003.

\bibitem{LBW99}
D.~A. Lidar, D.~Bacon, and K.~B. Whaley, ``Concatenating decoherence-free
  subspaces with quantum error correcting codes,'' {\em Phys. Rev. Lett.},
  vol.~82, pp.~4556--4559, May 1999.

\bibitem{Rus00}
M.~B. Ruskai, ``{Pauli Exchange Errors in Quantum Computation},'' {\em Phys.
  Rev. Lett.}, vol.~85, pp.~194--197, July 2000.

\bibitem{PoR04}
H.~Pollatsek and M.~B. Ruskai, ``{Permutationally invariant codes for quantum
  error correction},'' {\em Linear Algebra and its Applications}, vol.~392,
  no.~0, pp.~255--288, 2004.

\bibitem{ouyang2014permutation}
Y.~Ouyang, ``Permutation-invariant quantum codes,'' {\em Physical Review A},
  vol.~90, no.~6, p.~062317, 2014.

\bibitem{Pax97}
V.~Paxson, ``End-to-end internet packet dynamics,'' {\em SIGCOMM Comput.
  Commun. Rev.}, vol.~27, pp.~139--152, Oct. 1997.

\bibitem{LNCY97}
D.~W. Leung, M.~A. Nielsen, I.~L. Chuang, and Y.~Yamamoto, ``{Approximate
  quantum error correction can lead to better codes},'' {\em Phys. Rev. A},
  vol.~56, p.~2567, 1997.

\bibitem{BaK02}
H.~Barnum and E.~Knill, ``{Reversing quantum dynamics with near-optimal quantum
  and classical fidelity},'' {\em Journal of Mathematical Physics}, vol.~43,
  p.~2097, Jan. 2002.

\bibitem{Fletcher08}
A.~S. Fletcher, P.~W. Shor, and M.~Z. Win, ``{Channel-Adapted Quantum Error
  Correction for the Amplitude Damping Channel},'' {\em IEEE Transactions on
  Information Theory}, vol.~54, pp.~5705--5718, Dec. 2008.

\bibitem{Yam09}
N.~Yamamoto, ``{Exact solution for the max-min quantum error recovery
  problem},'' in {\em Decision and Control, 2009 held jointly with the 2009
  28th Chinese Control Conference. CDC/CCC 2009. Proceedings of the 48th IEEE
  Conference on}, pp.~1433--1438, IEEE, Dec. 2009.

\bibitem{Tys10}
J.~Tyson, ``{Two-sided bounds on minimum-error quantum measurement, on the
  reversibility of quantum dynamics, and on maximum overlap using directional
  iterates},'' {\em Journal of Mathematical Physics}, vol.~51, p.~92204, June
  2010.

\bibitem{BeO10}
C.~B\'{e}ny and O.~Oreshkov, ``{General Conditions for Approximate Quantum
  Error Correction and Near-Optimal Recovery Channels},'' {\em Phys. Rev.
  Lett.}, vol.~104, p.~120501, Mar. 2010.

\bibitem{BeO11}
C.~B\'{e}ny and O.~Oreshkov, ``{Approximate simulation of quantum channels},''
  {\em Phys. Rev. A}, vol.~84, p.~022333, Aug. 2011.

\bibitem{ouyang2013truncated}
Y.~Ouyang and W.~H. Ng, ``Truncated quantum channel representations for coupled
  harmonic oscillators,'' {\em Journal of Physics A: Mathematical and
  Theoretical}, vol.~46, no.~20, p.~205301, 2013.

\bibitem{BGu13}
M.~Bergmann and O.~G\"{u}hne, ``{Entanglement criteria for Dicke states},''
  {\em Journal of Physics A: Mathematical and Theoretical}, vol.~46, no.~38,
  p.~385304, 2013.

\bibitem{MHT12}
T.~Moroder, P.~Hyllus, G.~T\'{o}th, C.~Schwemmer, A.~Niggebaum, S.~Gaile,
  O.~G\"{u}hne, and H.~Weinfurter, ``{Permutationally invariant state
  reconstruction},'' {\em New Journal of Physics}, vol.~14, no.~10, p.~105001,
  2012.

\bibitem{TGG09}
G.~T\'{o}th and O.~G\"{u}hne, ``{Entanglement and Permutational Symmetry},''
  {\em Phys. Rev. Lett.}, vol.~102, p.~170503, May 2009.

\bibitem{zhang2014bounded}
Y.~Zhang, ``Bounded gaps between primes,'' {\em Annals of Mathematics},
  vol.~179, no.~3, pp.~1121--1174, 2014.

\bibitem{maynard2013small}
J.~Maynard, ``Small gaps between primes,'' {\em arXiv preprint
  arXiv:1311.4600}, 2013.

\bibitem{PBM86}
A.~Prudnikov, Y.~A. Brychkov, and O.~Marichev, {\em Integrals and Series,
  {V}olume 1, Elementary Functions}.
\newblock Gordon and Breach, 1986.

\bibitem{BuB05PRA}
D.~Burgarth and S.~Bose, ``Conclusive and arbitrarily perfect quantum-state
  transfer using parallel spin-chain channels,'' {\em Phys. Rev. A}, vol.~71,
  p.~052315, May 2005.

\bibitem{BGB05}
D.~Burgarth, V.~Giovannetti, and S.~Bose, ``Efficient and perfect state
  transfer in quantum chains,'' {\em Journal of Physics A: Mathematical and
  General}, vol.~38, no.~30, p.~6793, 2005.

\bibitem{BuB05NJP}
D.~Burgarth and S.~Bose, ``Perfect quantum state transfer with randomly coupled
  quantum chains,'' {\em New Journal of Physics}, vol.~7, no.~1, p.~135, 2005.

\bibitem{SJBB07}
K.~Shizume, K.~Jacobs, D.~Burgarth, and S.~Bose, ``Quantum communication via a
  continuously monitored dual spin chain,'' {\em Phys. Rev. A}, vol.~75,
  p.~062328, Jun 2007.

\bibitem{FV}
J.~Fitzsimons and T.~Vidick, ``A multiprover interactive proof system for the
  local hamiltonian problem,'' in {\em Proceedings of the 2015 Conference on
  Innovations in Theoretical Computer Science}, pp.~103--112, ACM, 2015.

\bibitem{Ji}
Z.~Ji, ``Classical verification of quantum proofs,'' {\em arXiv preprint
  arXiv:1505.07432}, 2015.

\end{thebibliography}

\bibliographystyle{ieeetr}
\end{document}